\documentclass[12pt, a4paper]{article}

\newcommand{\be}{\begin{equation}}
\newcommand{\ee}{\end{equation}}

\usepackage{graphicx}
\usepackage{amsmath}

\usepackage{a4}
\usepackage{amssymb}
\usepackage{subfigure}
\usepackage{color}
\usepackage[bf,footnotesize]{caption}

\def\glasg{a}
\def\liv{b}
\def\hum{c}

\begin{document}

\begin{titlepage}
  {\vspace{-0.5cm} \normalsize
  \hfill \parbox{60mm}{LTH823\\HU-EP-09/09\\SFB/CPP-09-23
                       }}\\[10mm]
  \begin{center}
    \begin{LARGE}
      \textbf{The  $\omega$-$\rho$  meson mass splitting and mixing
      from lattice QCD.}\\

    \end{LARGE}
  \end{center}

  \vskip 0.5cm
  \begin{figure}[h]
    \begin{center}
      \includegraphics[draft=false]{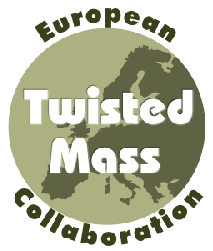}
    \end{center}
  \end{figure}

  \vspace{-0.8cm}
  \baselineskip 20pt plus 2pt minus 2pt
  \begin{center}
    \textbf{
      C.~McNeile$^{\glasg)}$,
      C.~Michael$^{(\liv)}$,
      C.~Urbach$^{(\hum)}$}\\
  \end{center}
  
  \begin{center}
    \begin{footnotesize}
      \noindent 
        $^{(\glasg)}$ 
 Department of Physics and Astronomy, The Kelvin Building,\\
 University of Glasgow, Glasgow G12 8QQ, UK\\
      \vspace{0.2cm}

      $^{(\liv)}$ Theoretical Physics Division, Dept. of Mathematical Sciences,
      \\University of Liverpool, Liverpool L69 7ZL, UK\\
      \vspace{0.2cm}

      $^{(\hum)}$ Institut f\"ur Elementarteilchenphysik, Fachbereich Physik,
\\ Humbolt Universit\"at zu Berlin, D-12489, Berlin, Germany\\      

    \end{footnotesize}
  \end{center}
  \vspace{2cm}

  \begin{abstract}
    \noindent\begin{tabular*}{1.\linewidth}{@{\extracolsep{\fill}}l}
      \hline
    \end{tabular*} 
    
    We compare  flavour singlet and non-singlet  vector mesons  from
first principles using lattice QCD. With $N_f=2$ flavours of light
quark, this addresses the $\omega$-$\rho$  mass difference.
 Using maximally twisted-mass lattice QCD, we  are able for the first time 
to determine this  mass difference precisely and we compare with experiment.
We also discuss $\omega$-$\rho$ mixing effects arising within QCD through 
the $u$-$d$ quark mass difference.

    \noindent\begin{tabular*}{1.\linewidth}{@{\extracolsep{\fill}}l}
      \hline
    \end{tabular*} 
  \end{abstract}

\end{titlepage}


\section{Introduction}

 The vector mesons are well described by a quark model since they are 
approximately `ideally' mixed, with the flavour non-singlet ($\rho$) and
 flavour singlet ($\omega$) degenerate. Experimentally ~\cite{pdg} the
$\omega$-$\rho$ mass  splitting is small (7 MeV). This is
 in contrast to the pseudoscalar mesons where the $\eta$ and $\eta'$ are
 much heavier than the $\pi$.  From first principles in QCD, this 
splitting of flavour singlet and non-singlet meson occurs because  of
contributions from disconnected quark diagrams. These disconnected 
contributions can be evaluated explicitly using lattice QCD. Early
lattice results showed that the disconnected diagrams are relatively
small for vector mesons~\cite{McNeile:2001cr}  but are relatively large
for pseudoscalar (and scalar) mesons.  Here we use more precise methods
to  determine the size of the disconnected diagrams for vector mesons
and we discuss the resulting phenomenology.

We describe  the vector mesons in a quark model basis for the case of 
degenerate $u$ and $d$ quarks. In the flavour singlet sector,  we then
have contributions  to the mass  matrix with quark model content $(u
\bar{u} +d \bar{d})/\sqrt{2}$  and $s \bar{s}$ (which we label as $nn$
and $ss$ respectively):
 \begin{equation}
  \label{eq:qmc}
  \begin{pmatrix}
    m_{nn} +2x_{nn} &  \sqrt{2} x_{ns} \\
                      & \\
    \sqrt{2} x_{ns}   &  m_{ss}+x_{ss}
  \end{pmatrix} \ .
\end{equation} 
 Here $m$ corresponds to the mass of the flavour non-singlet  eigenstate
and is  the contribution to the mass coming from connected fermion
diagrams while $x$ corresponds to the contribution from disconnected
fermion diagrams.  Thus $m_{nn}$ is the $\rho$ mass. Because of mixing, 
as will be discussed, $m_{ss}$ does not correspond exactly to any
specific meson. Since the $K^*$ meson has no disconnected contribution, 
one approximation is to use the assumption that  the connected
contribution to the meson mass  is linear in the underlying quark
masses. Then $m_{ss}= 2 m_{ns}-m_{nn}$,  that is $
2m_{K^*}-m_\rho$, leading to $m_{ss}=1.012$ GeV.

The mixing between the $nn$ and $ss$ flavour singlet channels must
produce the  experimental $\omega(782)$ and $\phi(1020)$. 
 We can express the mass  eigenstates as 
 \be
 m_{nn}+2 x_{nn} - \delta ; \ \ m_{ss}+x_{ss} + \delta
 \ee
 where $\delta=2x_{ns}^2/(m_{ss}+x_{ss}-m_{nn}-2x_{nn})$ to a good
approximation  for the relevant parameter values.  The $\delta$ term 
arising from mixing will prove to be  small (around 1 MeV mass shift) so
that one can estimate the $\omega$-$\rho$ mass difference as $2x_{nn}$ 
and the upward shift of the $\phi$ mass due to disconnected diagrams as
$ x_{ss}$. Even though the $\omega$ - $\phi$ mixing induces rather small
mass shifts,  the amplitude mixing can have significant effects. For
instance the physical  $\phi$ meson will have  a relative amplitude of
$\bar{n}n$ quarks (compared to $\bar{s}s$; this ratio is $\tan \delta$ where 
$\delta$ is the $\omega - \phi$ mixing  angle)
 of $\sqrt{2}x_{ns}/(m_{ss}+x_{ss}-m_{nn}-2x_{nn})$ which will be of
order a few percent and  can have an important consequence in $\phi$
decays to non-strange final states and $\omega$ production from strange 
initial states.

The contributions $x$ from disconnected diagrams will depend on the
quark  mass. For pseudoscalar mesons, the $x$ values are rather constant
at small  quark masses~\cite{Michael:2007vn} and decrease slowly with
increasing  quark mass~\cite{McNeile:2000hf}. Something similar would be
expected  for the vector mesons. 
 One simple phenomenological ansatz would be to assume all $x$-values 
were the same. Then the $\omega$-$\rho$ mass difference ($2x_{nn}=7.2$ MeV
experimentally) should be  approximately twice the $\phi$ mass shift
($x_{ss}$) although with the above model for  $m_{ss}$ this is  7 MeV
which does not agree.
 Such quantitative comparisons are not to be trusted for several reasons:
 (i) the $\rho$ meson is so wide (circa 150 MeV) that the impact of its
decay on its mass  value must introduce an  uncertainty of a few MeV at
least (ii) the model to estimate  the mass shift of the `connected'
$\phi$ (meson mass  linear in valence quark  content) is also imprecise
(for example, a meson mass-squared linear in valence quark content gives
5 MeV instead) and (iii) the underlying assumption  that the disconnected
contributions ($x$) are independent of mass may be at fault.
 To clarify these ideas, a direct evaluation  from QCD is necessary.

In this paper,  we study lattice QCD in the unitary sector with $N_f=2$
degenerate light quarks. This enables us to extract estimates for the
$\omega$-$\rho$ mass difference. We can also explore the (sea)quark mass
dependence  of $x_{nn}$ which can help to understand the value of
$x_{ss}$ needed  to determine the disconnected contribution to the
$\phi$ meson.

There is considerable interest in building reliable  models of the quark
mass dependence of the $\rho$ meson. We show that such models inevitably
have consequences for  the $\omega-\rho$ mass difference. We compare
these predictions with our  lattice results.

We also explore effects arising in QCD from the $u$-$d$ quark mass
difference.  This causes a violation of isospin and a mixing between
$\omega$ and $\rho$  mesons. We are able to evaluate this mixing (given
the quark mass difference as  input) from the lattice and we compare
with experiment.

Here we follow the lattice methods used in our study of the flavour 
singlet pseudoscalar mesons~\cite{Jansen:2008wv}. The disconnected
contributions  are evaluated using stochastic methods with variance
reduction while the  connected contributions are evaluated using
stochastic time-plane sources with the `one end trick'. 

 Use of the remarkable precision
obtainable~\cite{Michael:2007vn,Boucaud:2008xu,Jansen:2008wv}  in
evaluating  disconnected contributions in maximally-twisted lattice QCD
will enable us to obtain a statistically significant signal  for the
vector mass splitting.

\section{Twisted mass lattice QCD and neutral particles}

In quenched or partially-quenched lattice QCD, the disconnected
contribution  to the flavour singlet meson does not combine properly
with the  connected contribution to give a physical state. To avoid this
problem,  it is mandatory to study flavour singlet states in full QCD -
with  sea quarks having the same properties as valence quarks. Then the
spectrum  of flavour singlet states is well defined and can be extracted
from the  $t$-dependence of the full correlator.  Here we focus on the
case where there  are two degenerate light quarks (called $N_f=2$) which
is a consistent  theory in which to study the flavour singlet mesons. We
use the twisted mass lattice
formalism~\cite{Frezzotti:2000nk,Frezzotti:2003ni}, for a recent review,
see ref.~\cite{Shindler:2007vp}. 

 The results presented in this paper are based on gauge configurations
as produced by the European Twisted Mass collaboration (ETMC). The
details of the ensembles are described in ref.~\cite{Urbach:2007rt}. In
particular, we concentrated for this paper on the ensembles labelled
$B_1, B_2, B_3, B_6$ and $C_1$ and  $C_2$. We have compiled the details for
those ensembles in table~\ref{tab:setup}.  The $B$-ensembles correspond
to a value of the lattice spacing of about $a\sim0.09\ \mathrm{fm}$ and
the $C$-ensembles to $a\sim0.07\ \mathrm{fm}$. The spatial lattice size
is of about $L\sim2.2\ \mathrm{fm}$ for all ensembles used here, apart
from $B_6$, which has identical parameters to $B_1$ but $L\sim2.7\
\mathrm{fm}$. The corresponding pseudoscalar mass is in the range 
300-440 MeV.

\begin{table}[t!]
  \centering
  \begin{tabular*}{0.9\textwidth}{@{\extracolsep{\fill}}lccccc}
    \hline\hline
    Ensemble & $L^3\times T$ & $\beta$ & $a\mu_q$ & $\kappa$ & $r_0/a$\\
    \hline\hline
    $B_1$ & $24^3\times 48$ & $3.9$ & $0.0040$ & $0.160856$ & $5.22(2)$\\
    $B_2$ &  & & $0.0064$ &  &\\
    $B_3$ &  & & $0.0085$ &  &\\
    $B_6$ & $32^3\times 64$ & $3.9$ & $0.0040$ & $0.160856$ &\\
    \hline
    $C_1$ & $32^3\times 64$ & $4.05$ & $0.003$ & $0.157010$ & $6.61(3)$\\
    $C_2$ & & & $0.006$ & &\\
    \hline\hline
  \end{tabular*}
  \caption{Summary of ensembles produced by the ETM collaboration used
    in this work. We
    give the lattice volume $L^3\times T$ and the values of the
    inverse coupling $\beta$, the twisted mass parameter $a\mu_q$, the
    hopping parameter $\kappa=(2am_0+8)^{-1}$ and the Sommer parameter
  $r_0/a$ in the chiral limit from ref.~\cite{Boucaud:2007uk,Urbach:2007rt}.
   The data sets cover 5000 equilibrated trajectories (10000 for $B_1$).
 }
  \label{tab:setup}
\end{table}

The twisted mass formalism at maximal twist is order $a$ improved 
and allows access to light pseudoscalar mesons~\cite{Boucaud:2007uk}.
There is one complication, however, namely that the twisted mass lattice 
formalism breaks flavour and parity symmetries at finite values of the
lattice spacing $a$. These symmetries are restored  in the continuum
limit and the theory is well defined (so providing a valid 
regularisation) at finite lattice spacing. One consequence of this
flavour-breaking is that  charged and neutral non-singlet mesons can have
a mass  splitting (of order $a^2$). Moreover, the neutral non-singlet
mesons  have (order $a^2$) contributions from disconnected
diagrams~\cite{Jansen:2005cg,Boucaud:2007uk}. Note that the  order $a^2$
splitting between $\rho^+$ and $\rho^0$ was found to be  compatible with
 zero~\cite{Michael:2007vn}.

In the case of twisted mass fermions, the bilinear operators appropriate
to  create the $\omega$ state are $O^V_i=\bar{\psi} \gamma_i \psi$ and
 $O^T_i=\bar{\psi} \gamma_i \gamma_4 \psi$ which  on  transformation
into the twisted basis (used in lattice evaluation) will become 
$\bar{\chi} \gamma_i \chi$ and $\bar{\chi} \gamma_i \gamma_4\gamma_5
\tau_3 \chi$, respectively, where $\tau_3$ acts in the $(u,\ d)$ flavour
space. This latter case amounts to evaluating, in the lattice basis, the
difference of the  disconnected loop between $u$ and $d$ quarks. 
 This enables a very  efficient variance reduction
technique~\cite{Michael:2007vn,Boucaud:2008xu,Jansen:2008wv} to be used to
evaluate the  relevant disconnected diagram for the case of the $O^T$ 
operator.
 Here we follow in detail the methods to evaluate the disconnected ($D$)
and connected ($C$)  meson correlators described in
ref~\cite{Jansen:2008wv}. Some  discussion of the connected correlators
for  vector mesons has already
appeared~\cite{dimopoulos:2008ee,dimopoulos:2008hb} and a fuller study 
is in preparation~\cite{etmcrho}.

 The connected meson correlator describes the flavour non-singlet meson
($\rho$),  while the combined correlator describes the flavour singlet
($\omega$). Assuming sufficiently large $t$-separation so that the 
ground states dominate, then we have (for $N_f=2$) 
 \be
  C(t)=c \exp(-m(\rho) t) ; \ \ C(t)-2 D(t)= d \exp(-m(\omega) t)
 \ee
 
 Hence 
 \be
  2 D(t) /C(t) = -{d \over c} \exp(-\Delta t) + 1
 \ee
 where $\Delta=m(\omega)-m(\rho)$. Moreover, if $\Delta$ is small, 
as we shall find, then 
 \be
  2 D(t) /C(t) = {d \over c} \Delta t + ( 1 - {d \over c})
 \label{equ:dcline}
 \ee
 Since the difference between $\rho$ and $\omega$ is so small, one would
also  expect the creation amplitudes to be equal so that $d \approx c$.
Then keeping  the first order in the mass splitting, this  leads to the
simple result that $2 D(t) /C(t) =  \Delta t + {\cal O}(\Delta)$ which will be a good
approximation  provided $\Delta t \ll 1$.

 In twisted mass QCD, there is an order $a^2$ contribution to the
$\rho^0$ meson propagation arising from disconnected diagrams. This has 
been studied previously~\cite{Michael:2007vn} using the operator $O^V$
which enables the  variance reduction method to give precise results in
this case. No statistically  significant contribution was found. For our
present purposes, however, we wish  to focus on the disconnected
contribution to the $\omega$ meson compared to its connected
contribution which is in turn equal to the connected contribution of the
$\rho^0$ propagation. So we compare  the $\omega$ propagation with that
of the connected contribution ($C$) for the $\rho^0$ meson.

\begin{figure}[t]
  \centering
  \includegraphics[width=.8\linewidth]{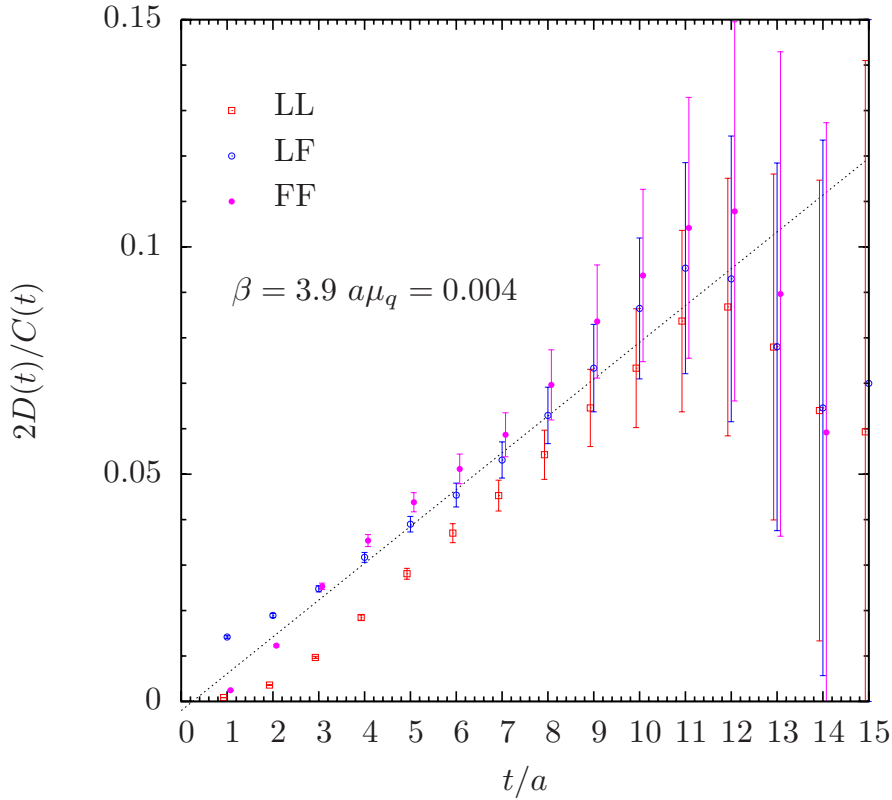} 
  
  \caption{Ratio of disconnected to connected contribution to neutral
vector  meson (created by $O^T$) versus $t$. Results are from ensemble
$B_1$ ($L=24$)  with $a\mu_q=0.004$. The ratio is given for cases
with local operator at source and sink (LL) and for cases when 
one or both operator is non-local (fuzzed F). The line represents the fit 
to the LF data in the $t/a$-range 5 to 12.
  }
  \label{fig:dbyc004}
\end{figure}

 We present some results for this ratio $2D/C$ in
fig.~\ref{fig:dbyc004}. We have evaluated this for local meson operators
(L) and non-local (fuzzed) meson sources (F).

\begin{figure}[t]
  \centering
  \includegraphics[width=.8\linewidth]{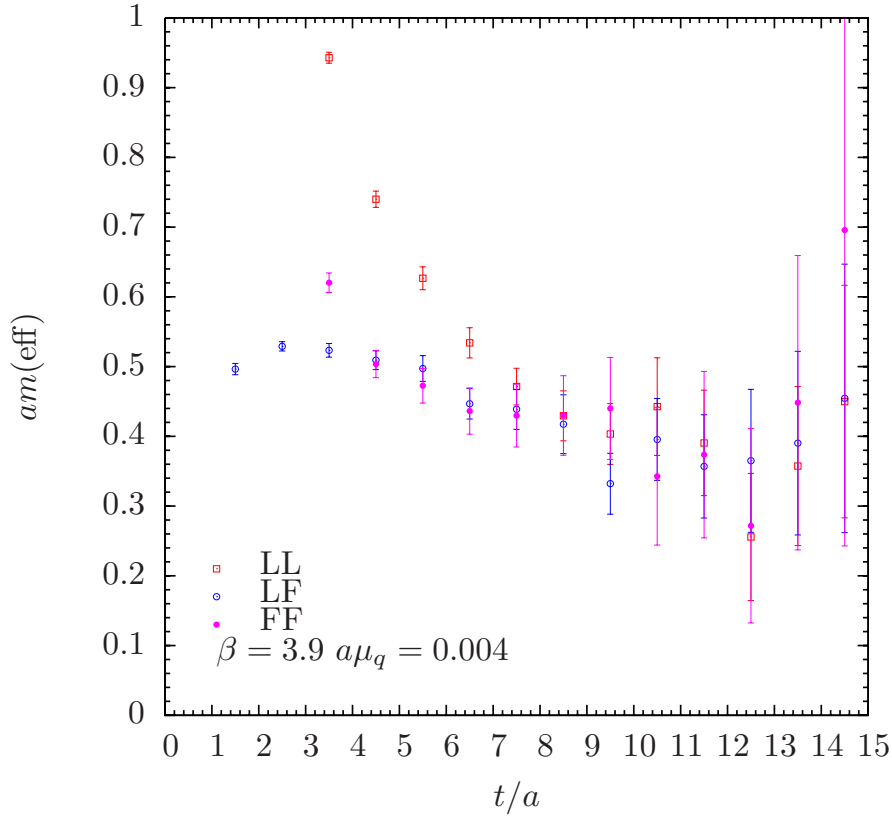} 
  
  \caption{The effective mass (in lattice units) of the connected
contribution to the neutral vector  meson (created by $O^T$) versus $t$.
Results are from ensemble $B_6$ ($L=32$)  with $a\mu_q=0.004$ and
$r_0/a=5.22$. The result is given for cases with local operator at source
and sink (LL) and for cases when  one or both operator is non-local
(fuzzed F).
  }
  \label{fig:meff}
\end{figure}

 In order to interpret this signal we evaluate the effective  mass  from
the connected component which is displayed in fig~\ref{fig:meff}. This
shows that the ground state dominates in this case for $t/a > 5$ for the
LF and  FF cases.   This allows us to interpret the data in
fig.~\ref{fig:dbyc004}  as being dominated by the ground state for that
$t$-region, so that one  can determine the mass splitting from the slope
as described above.

We fit the behaviour of $2D/C$ as a straight line 
(eq.~\ref{equ:dcline}) in the range of  $t/a$ from 5 to 12 (6 to 12 for
the LL results) for $\beta=3.9$ and 6 to 16 (7 to 15 for LL) for
$\beta=4.05$. The $\chi^2$ of these  fits is acceptable. We take the
average of the slope from the LF and FF fits and we  determine an error 
for the slope $a \Delta$ from combining the range of values found
(including the LL case). The intercepts are very close to zero, so the
impact of the factor $d/c$ on  the determination of $\Delta$ in
eq.~\ref{equ:dcline} is negligible. We find that the results from two
different volumes  (spatial extent L=24 and 32 at $\mu_q=0.004$) are
consistent with each other.

\begin{figure}[t]
  \centering
  \includegraphics[width=.8\linewidth]{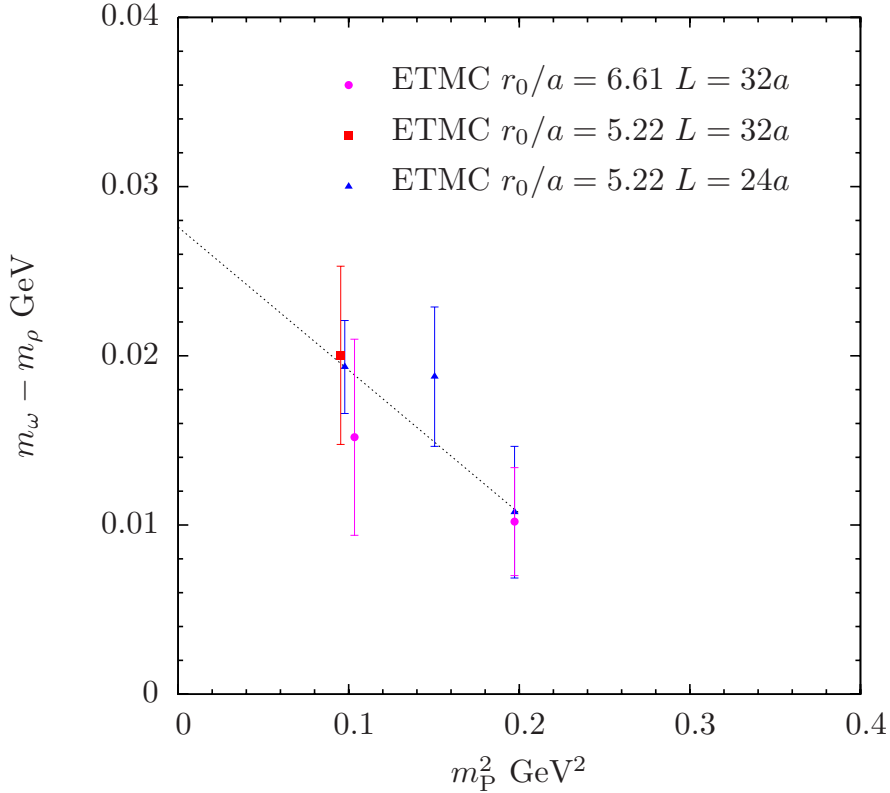} 
  
  \caption{The $\omega-\rho$ mass difference  versus the quark mass
given by $ m_{P}^2$.    A linear extrapolation is plotted.
 The lattice spacing is $a$  and the  spatial extent of the lattice is
$L$.
  }
  \label{fig:all}
\end{figure}

 In order to compare the  results from different lattice
spacings, we plot them using the $r_0/a$-values given in
table~\ref{tab:setup} to convert our results to GeV units 
using $r_0=0.454(7)$fm obtained~\cite{Boucaud:2007uk} from  the ETMC
evaluation of $f_{\pi}$. 
  We show the  $\omega$-$\rho$ mass splitting in fig.~\ref{fig:all} and
we see that our results at different lattice spacings are consistent
with each other.

 Extrapolation to the physical limit (for $u$ and $d$ quarks)
 gives $m(\omega)-m(\rho) =27(5)$ MeV assuming a  linear dependence.
Since there is  limited evidence for the form of the extrapolation,  we
assign a error which covers the possibility of a constant  extrapolation
(namely 27(10) MeV).

 This may be compared with the experimental splitting~\cite{pdg} of 7.2
MeV.
 In our QCD study, we do not include electromagnetic effects or effects 
arising from the $u$-$d$ quark mass difference. These effects are 
estimated~\cite{Bijnens:1996kg} to be of the order of a few MeV.
Furthermore, the definition of the  $\rho$ mass is uncertain to a
similar extent, because it is such a  wide resonance.  Hence precise
agreement between the observed $\omega$-$\rho$  mass difference and that
found from lattice QCD is not to be expected.

 The quark mass dependence of vector meson masses is of considerable 
interest, especially in the lattice community. Next we 
discus models which may be compared with our results.

\section{Models for vector meson masses}

The disconnected quark diagram that contributes to the  $\omega-\rho$
mass difference  can be used to inform models of the mass-dependence of
the vector  meson masses on the underlying quark mass.  Such models 
contain contributions from two-particle intermediate states evaluated 
in some effective theory.   One contribution to the $\omega-\rho$
mass difference comes from  such two particle contributions to the self
energy. This gives an  excellent way to check on these effective Lagrangian
models for  the $\rho$ mass versus $m_{\pi}$. Here we retain the physical 
meson names ($\rho$, $\omega$, $\pi$) to describe mesons with these 
quantum numbers as the light quark mass is varied.

\begin{figure}[t]
  \centering
  \includegraphics[width=.3\linewidth]{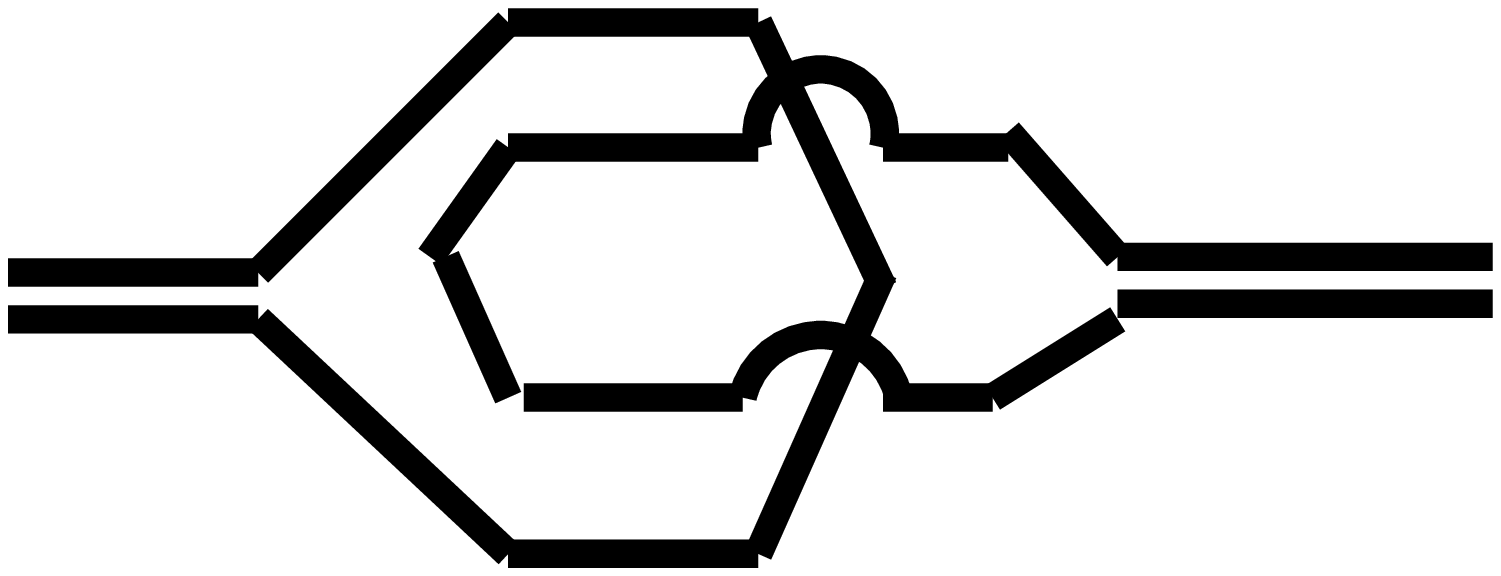} \ \ \ \ \ 
  \includegraphics[width=.3\linewidth]{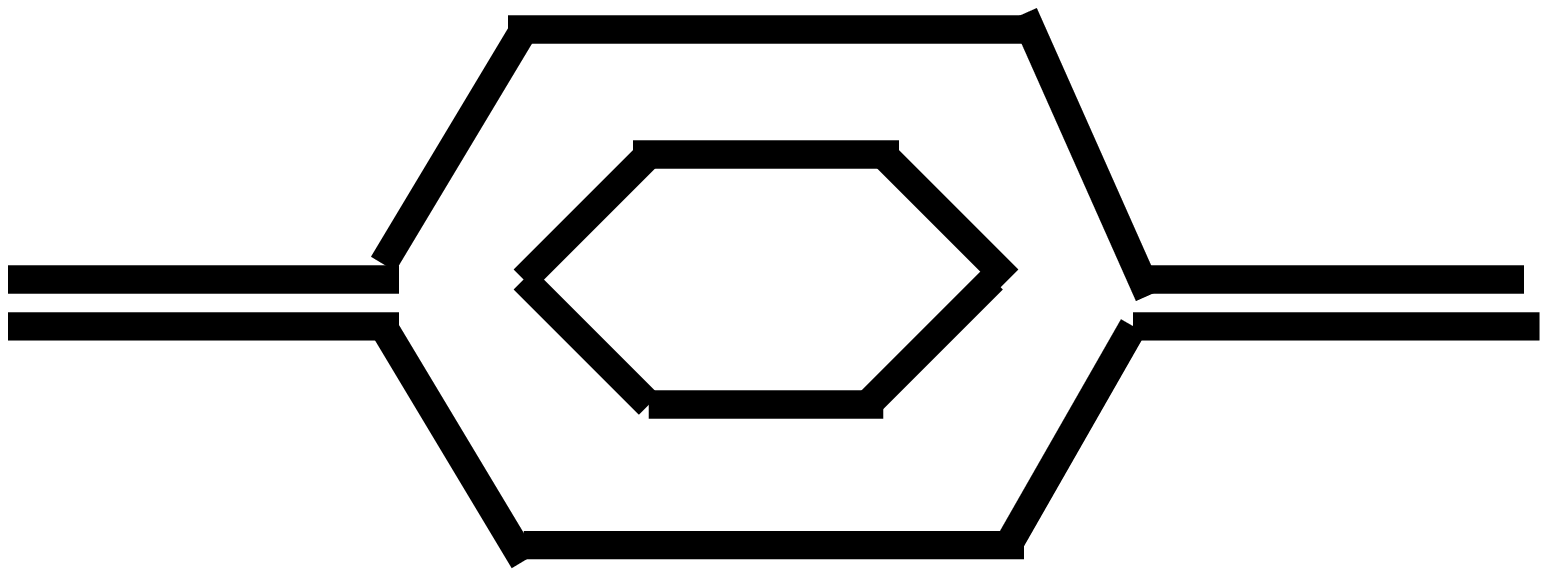}   

  \caption{
Disconnected and connected two-particle intermediate states.
  \label{fig:disgraph}
  }

\end{figure}

One way to visualise this contribution to disconnected diagrams from 
two-particle intermediate states is illustrated by  the two quark diagrams 
describing disconnected and connected quark line contributions in
fig.~\ref{fig:disgraph}.

The  two-body contributions (having non-analytic behaviour with $m_{\pi}$) 
that are usually considered are the pseudoscalar-pseudoscalar (PP) and
vector-pseudoscalar (VP) intermediate states. Here we tabulate the relative 
contributions to $\rho$ and $\omega$ mesons:

\begin{tabular}{l l l}
      V  &  PP  & VP \\
     $\rho$: &   $\pi\ \pi$  & $\omega \pi$ (1$g^2$) \\
   $\omega$: &    -      &     $\rho \pi$ (3$g^2$) \\
\end{tabular}

 The notation for the VP case indicates that  $\omega \to  \rho \pi$ has
3 terms ($\rho^0 \pi^0$,  $\rho^+ \pi^-$, $\rho^- \pi^+$) whereas    $\rho
 \to \omega \pi$ has just one ($\omega^0 \pi^0$ but with the same
coupling).
 Thus the contributions from these intermediate states 
to the $\omega-\rho$ mass difference are  $  VP(2g^2) - PP $  while
the $\rho$ mass has              $  VP(1g^2) + PP$.
 This exemplifies the link between models for the quark mass dependence 
of the $\rho$ meson and models for the $\omega-\rho$ mass difference.

 We now discuss some models proposed and focus on their implications for
the  $\omega-\rho$ mass difference. We first consider the VP
intermediate state. The behaviour of this contribution versus quark mass
is as $m_{\pi}^3$ and  estimates of the contribution (without using any
form factor) to the $\rho$ mass at the physical pion mass  are  3.4
MeV~\cite{Bijnens:1996kg} and 5 MeV~\cite{Leinweber:2001ac}. The strong
dependence on $m_{\pi}$ implies  that, in these models, the VP
contribution for a  pion mass around 300 MeV will  be much bigger: for
the $\omega-\rho$ mass difference they obtain values of  -70
MeV~\cite{Bijnens:1996kg} and -30 MeV~\cite{Leinweber:2001ac} using a
form factor cutoff. This large value obtained without cutoff is quoted
as `not to be taken  too seriously' by ref.~\cite{Bijnens:1996kg}. For
more discussion of  regulators which may improve the convergence, see
also ref.~\cite{Bruns:2004tj}.  In our opinion, the evaluation of the VP
intermediate state is delicate since as the quark mass decreases,  the V
and VP states become nearly degenerate. In this circumstance, one needs
to take account of the width of the  $\rho$ meson which could modify the
expressions significantly.

The PP intermediate state (i.e. $\pi \pi$ contribution to the $\rho$)
has  a non-analytic component behaving as $m_{\pi}^4 \log (m_{\pi})$.
One  approach~\cite{Leinweber:2001ac} uses the relativistic matrix
element but introduces a form factor to emphasize the kinematic region 
where the model should be more reliable. Fitting to lattice data, they 
obtain a contribution to the $\rho$ mass of -35 MeV for the physical
pion mass  and -40 MeV for pion masses around 300 MeV. These values
imply a contribution  to the $\omega-\rho$ mass difference of 40 MeV
for  pion masses around 300 MeV.

In summary, models give quite large contributions to the $\omega-\rho$
mass difference, with the VP intermediate state dominating, and becoming
 more negative as the quark mass is increased. However the models  have
an increasingly large contribution at larger masses  which needs to be
tamed by a cutoff or other method.
 Our lattice data  show  a decrease with increasing quark mass, but with
numerical values that  are much smaller than those typically  given by
models.

\section{$\omega-\rho$ mixing}

 Within QCD when $m_u \ne m_d$, isospin is not a good quantum number and
there is a mixing matrix element between the $\omega$ and $\rho$ mesons.
 One experimental manifestation of this is the decay $\omega \to \pi
\pi$ which  creates an interference pattern in the dominant decay $\rho
\to \pi \pi$  near the $\omega$ mass. The mixing can be defined from a
mass matrix (with  basis states $(\bar{u} \gamma_i u \pm \bar{d}
\gamma_i d)/\sqrt{2}$ ):

 \begin{equation}
  \label{eq:om-rho}
  \begin{pmatrix}
    m_{\rho}  &   T_{\omega \rho} \\
                      & \\
    T_{\omega \rho}   &  m_{\omega} 
  \end{pmatrix} \ .
\end{equation} 

\noindent  Phenomenologically~\cite{Miller:2006tv} the transition matrix
element  $T_{\omega \rho} = -3.1(3)$ MeV. Some of the observed effect
can come from  electromagnetic contributions and these are
estimated~\cite{Bijnens:1996kg} to contribute  between 0.4 and  1.2 MeV.
So the QCD contribution would be around -4 MeV.

We can study this contribution on a lattice by considering the  cross
correlator: create a $\omega$ meson (using $(\bar{u} \gamma_i u+\bar{d}
\gamma_i d)/\sqrt{2}$) and destroy a $\rho$ meson (using $(\bar{u}
\gamma_i u-\bar{d} \gamma_i d)/\sqrt{2}$ ). This cross correlator will
have a connected  piece (like $(C_{u}-C_{d})/2$) and a disconnected
piece.  The  disconnected contribution simplifies using  the identity  
$D_{ud}=D_{du}$ yielding  $(D_{uu} -D_{dd})/2$. Here $D_{ud}$ means the
correlator between  a $u$-quark loop at source and a $d$-quark loop at
sink. Because of the form of the cross-correlator, it may be expressed 
as a difference between propagation of vector mesons containing  quarks
of different mass. It may then be related (using techniques similar to
those used in ref.~\cite{Foster:1998vw} and assuming that
$(m_{\omega}-m_{\rho})t \ll 1 $ to simplify the expression) to the 
mass-matrix transition element:

 \be
   T_{\omega \rho} = ({dm_{\rho} \over dm_q} + {dm_{\omega} \over dm_q})
 \    {m_u-m_d \over 4}
 \label{equ:mix}
 \ee


\noindent  where we have assumed that we may retain the first power of
$m_u-m_d$ only. Here  the derivatives are with respect to the 
degenerate quark mass in an  $N_f=2$ theory. These derivatives are those
that we have access to from our  study above.

 We can estimate this transition rate phenomenologically, using  $
dm_{\rho}/dm_{\pi}^2 \approx 1/(2 m_{\rho})$ (from the $\phi$, $K^*$ and
$\rho$ masses, see ref.~\cite{Lacock:1995tq}) and $dm_{\pi}^2/dm_q = 
m_{\pi}^2/\hat{m_q}$ (from lowest order ChPT). Here $\hat{m_q}$ is the
average light quark mass $(m_u+m_d)/2$. This gives us 

 \be
   T_{\omega \rho}= -{1-m_u/m_d \over 1+m_u/m_d} \ {m_{\pi}^2 \over 2 m_{\rho}}
 \ee
 and with $m_u/m_d=0.56$ from lowest order of ChPT~\cite{pdg} we obtain
 $ T_{\omega \rho}= -3.6$ MeV which is close to the experimentally determined 
value.

 We now discuss evaluation of the above expression in eq.~\ref{equ:mix}
using  lattice results from first principles. This involves the quark
mass dependence  of the $\rho$ and $\omega$ masses in an $N_f=2$ theory,
which we have available. In particular we are able to address the quark mass 
dependence of the disconnected contribution to the $\omega$ mass, which has 
not been studied previously.

Since ${dm_{\rho} / dm_q} + {dm_{\omega} / dm_q}= 2dm_{\rho} /d m_q +
d(m_{\omega}-m_{\rho}) /d m_q $ we compare these two terms.
 We can evaluate
 $ d (m_{\omega}-m_{\rho}) / d m_P^2 $ as the slope shown in
fig.~\ref{fig:all}.  We obtain a slope $-0.15(15)/(2m_V)$ where the error
encompasses the  possibility of a constant behaviour which is not
strongly excluded by our lattice results.  Compared to the
phenomenological estimate that  $ dm_{\rho} /dm_P^2 \approx 1/(2m_V)$,  this
value indicates that the  contribution of the $\omega$ meson
disconnected diagram to the  $\omega - \rho$ mixing is relatively
insignificant (at around 7\% )- in agreement with  earlier phenomenological
estimates~\cite{Bijnens:1996kg}.

We find that lattice QCD naturally produces effects of the correct 
size to explain the observed $\omega$-$\rho$ mixing.

\section{Summary}


Previous lattice results  were very exploratory: for $N_f=2$ sea quarks 
of mass (given by $(r_0 m_{PS})^2=3.7$) corresponding to  about 1.5 times
the strange quark mass, an $\omega$-$\rho$ mass difference of 2(3) MeV 
was obtained ~\cite{McNeile:2001cr} with quite strong evidence that  the
sign of the effect should be positive (the same as found here). A more
recent study finds~\cite{Hashimoto:2008xg}, in the chiral limit, an
$\omega$ mass  of 790(194) MeV, fixing the scale at the $\rho$ mass: so 
a mass difference of  15(194) MeV.

Here we have presented results  coming from a well established signal
and  we are able to address issues such as  extrapolation to the
continuum limit.  We see in fig.~\ref{fig:all}  that our results are
consistent with each other at different lattice spacing  and for
different lattice spatial volumes.
 There is  evidence for a reduction in the splitting with increasing 
quark mass. Since there is  limited evidence for the form of the
extrapolation to the  physical point,  we assign a error which
covers the possibility of a constant  extrapolation (namely 27(10) MeV).

 This may be compared with the experimental splitting~\cite{pdg} of 7.2
MeV.
 The experimental result includes electromagnetic effects  (estimated to
be of order a few MeV) and  the definition of the  $\rho$ mass is
uncertain to a similar extent, because it is such a  wide resonance. 
Hence precise agreement between the observed $\omega$-$\rho$  mass
difference and that found from lattice QCD is not to be expected. What 
we do establish, however, is that there are substantial contributions to
the  $\omega$-$\rho$ mass difference arising within QCD from the
disconnected contributions (known phenomenologically as OZI violating
contributions).  Moreover, these contributions are of the same sign as
the experimentally observed difference.

Our study shows that it is possible to extract a signal for the $\omega-\rho$ 
mass splitting versus quark mass and that this quantity will be very 
useful in refining models for vector mesons near the chiral limit. Indeed 
some existing models give results which are significantly different from 
those we obtain (especially for the VP intermediate state contribution).

We also showed from first principles that the observed $\omega$-$\rho$
mixing arises naturally  in QCD with the correct magnitude. We were 
able to estimate, for the first time, the contribution to this mixing
arising from  disconnected diagram contributions and we showed that it is
relatively  unimportant.

It has been suggested that the $\omega-\rho$ mixing is the mechanism
behind charge symmetry breaking in nucleon-nucleon interactions. However
there are other competing mechanisms (see Miller et
al.~\cite{Miller:2006tv} for a review) that try to explain charge
symmetry breaking. One problem with the proposed mechanism of
$\omega-\rho$ mixing for charge symmetry breaking in the nuclear force
is that the momentum dependence of the mixing was hard to constrain.
Although we have only evaluated the $\omega-\rho$ mixing at zero
momentum, our techniques can in principle be used to investigate the
momentum dependence~\cite{McNeile:2002az}, so our calculation is the
first step in determining the mechanism of charge symmetry breaking in
the nuclear force from first principles.

Although we have studied QCD with  $N_f=2$ light degenerate fermions,
one can extend the discussion to include the strange quark by  assuming
that the disconnected contribution is independent of the quark mass, so
yielding an estimate of $x_{ns} \approx  x_{nn}=13(5) $
MeV. This  implies that the $\phi$-$\omega$ Hamiltonian mixing  amplitude
 given by $\sqrt{2} x_{ns}$ should be of comparable size. This is  in
qualitative agreement with phenomenology~\cite{Bijnens:1996kg} which finds 
values of 8 to 14 MeV from an analysis of $\phi$ decays.

 The mixing between the $\omega$ and $\phi$ is required to understand $B$
and $D_s$ decays with an $\omega$ or $\phi$ in the final
state~\cite{Gronau:2008kk,Gronau:2009mp}. For example Gronau and
Rosner~\cite{Gronau:2009mp} predict the branching ratio 
 ${\cal B}( D_s^+ \rightarrow \omega e^+ \nu_e) / {\cal B}( D_s^+
\rightarrow \phi e^+ \nu_e) $ 
 in terms of the $\omega-\phi$ mixing angle. Other mechanisms can
contribute to  this ratio, such as "weak annihilation", so a first
principles calculation of the $\omega-\phi$ mixing angle is valuable.


In summary, the vector mesons have a rich structure beyond  `ideal
mixing' and this rich structure can be evaluated accurately using 
lattice QCD.

\subsubsection*{Acknowledgments}

 We acknowledge helpful advice from members of the ETM Collaboration and
 computing resources provided by NW Grid at Liverpool and by HLRN at
Berlin.


\end{document}